%% file: main.tex
\documentclass[aps,prl,twocolumn,superscriptaddress,nofootinbib]{revtex4-2}

\usepackage[english]{babel}
\usepackage[utf8]{inputenc}
\usepackage[T1]{fontenc}

\usepackage{amsmath,amssymb}
\usepackage{graphicx}
\usepackage{tikz}
\usetikzlibrary{quantikz2,positioning}
\usetikzlibrary{arrows.meta, positioning, shapes.geometric}
\usepackage{longtable}
\usepackage{booktabs}
\usepackage{adjustbox}
\usepackage{subfigure}
\usepackage{tabularx}
\usepackage[colorlinks=true, allcolors=blue]{hyperref}
\usepackage{graphicx}
\usepackage{dcolumn}
\usepackage{bm}
\usepackage{booktabs} 

\usepackage{amsmath} 
\usepackage{amsfonts} 
\usepackage{amssymb} 
\usepackage{algorithm} 
\usepackage{algpseudocode} 

\begin{document}

\title{Application of Quantum Annealing to Computation of Molecular Properties} 

\author{Pradyot Pritam Sahoo}
\email{pradyotpritam@gmail.com}
\affiliation{Centre for Quantum Engineering, Research and Education, TCG Crest, Kolkata 700091, India}
\affiliation{Academy of Scientific and Innovative Research (AcSIR), Ghaziabad 201002, India}

\author{V. S. Prasannaa }
\email{srinivasaprasannaa@gmail.com}
\affiliation{Centre for Quantum Engineering, Research and Education, TCG Crest, Kolkata 700091, India}
\affiliation{Academy of Scientific and Innovative Research (AcSIR), Ghaziabad 201002, India}

\author{B. P. Das}
\email{bpdas.iia@gmail.com}
\affiliation{Centre for Quantum Engineering, Research and Education, TCG Crest, Kolkata 700091, India}
\affiliation{Department of Physics, Institute of Science, Tokyo (Formerly Tokyo Institute of Technology), Ookayama, Meguro-ku, 152-8550, Japan
}

\date{\today}

\begin{abstract}

We present the results of our quantum annealing computations of the permanent electric dipole moments of several molecules. By 
applying an electric field as a perturbation and measuring the corresponding energy responses, the 
molecular electric dipole moments are obtained numerically through the finite field method. The ground-state electronic wavefunctions and 
energies are obtained using the quantum annealer eignsolver algorithm. This work provides a  pathway for the 
computation of molecular properties in the quantum annealing paradigm.

\end{abstract}

\maketitle


\textbf{}
Quantum annealing (QA) is a heuristic method for solving combinatorial optimization problems that leverages quantum effects such as tunneling and superposition~\cite{Kadowaki1998}. It represents a specialized approach closely related to Adiabatic Quantum Computing, designed to address computationally hard problems by harnessing these quantum effects~\cite{McGeoch2014}.
The Quantum Annealer Eigensolver (QAE), a quantum-classical hybrid 
algorithm based on the
variational principle~\cite{Sakurai2010}, has been demonstrated to yield highly accurate results 
for ground-state energy
optimization~\cite{vikrant2024}. Accurate calculations of physical properties is important in 
physics and chemistry as they provide insights to a wide range of applications in
molecules and materials. The QAE is broadly applicable to a wide range of 
physical optimization problems, 
including molecular vibrational spectra~\cite{Teplukhin2019}, complex eigenvalue problems, energy calculations for 
molecular electronic states, relativistic fine-structure splitting in highly charged atomic ions~\cite{vikrant2024}, 
particle physics simulations~\cite{Illa2022}, and lattice gauge theories~\cite{ARahman2021}.

In this work, we extend the application of  QA to compute an important molecular property- the 
permanent electric dipole moment (PDM). This study underscores the broad applicability of the method to molecules, thereby opening a new avenue for computing molecular properties beyond the ground-state energy. We compute the PDM of molecules 
adapting the finite-field method (FFM)  to  QA. Moreover, PDMs are important for various applications, 
particularly for SrF~\cite{pra2014}, the first molecule to be laser-cooled~\cite{Shuman2010} and and BaF, a promising candidate for precision measurements in fundamental physics~\cite{Aggarwal_2018}.

This paper begins with an overview of the FFM~\cite{pra2018} and the Dirac-Coulomb (DC) framework \cite{Grant2007}, which form the 
theoretical basis for relativistic calculations using quantum annealing. We then describe the methodology used in the 
Dirac–Fock (DF) calculations for the BeF, MgF, CaF, SrF, and BaF molecules. Following this, we introduce the QAE algorithm 
and explain its implementation on the D-Wave quantum annealer. Subsequently, the overall 
workflow of the problem is 
outlined. A comparative analysis of the results obtained from simulated annealing, quantum 
annealing, and full  configuration interaction (FCI)/exact diagonalization (ED) is subsequently presented. The paper concludes with a summary of our key findings and 
a discussion of potential directions for future research.

The PDM of a molecule is a fundamental 
property that quantifies the separation of positive and negative charges. 
The FFM offers a numerical technique to evaluate the PDM by measuring the variation of the molecular energy as a response to an 
external electric field~\cite{pra2018,Szabo:1996}. When an external electric field $\vec{E}$ is applied along the $z$-direction, the molecular Hamiltonian is perturbed as $\hat{H} = \hat{H}_0 + \epsilon \hat{O}$
where $\hat{H}_0$ is the unperturbed molecular Hamiltonian and $\hat{O}$ is the dipole moment operator in the $z$-direction, and $\epsilon$ is the perturbation strength.
From first-order perturbation theory, the energy correction due to the field is given by: $E(\epsilon) = E_0 + \epsilon \bra{\Psi_0} \hat{O} \ket{\Psi_0},$
where $E_0$ is the unperturbed energy and $\ket{\Psi_0}$ is the unperturbed wavefunction.
Thus, the dipole moment in the $z$-direction can be expressed as
$\langle\hat{O}\rangle =  \left| \frac{\partial E}{\partial \epsilon} \right|_{\epsilon \to 0} $,
instead of computing the derivative analytically, we approximate it numerically using the central difference formula:
\begin{equation}\label{eq: ffm}
    \langle\hat{O}\rangle \approx \lim_{\epsilon\to 0}\frac{E(+\epsilon) - E(-\epsilon)}{2\epsilon}.
\end{equation}
A small perturbative electric field --typically on the order of $10^{-4}$ to $10^{-3}$ a.u. is applied along the $z$-axis, and the total energy is computed for both positive and negative field values.  Both $E(+\epsilon)$ and $E(-\epsilon)$ are computed using the QAE.
The Hamiltonian  in this work is the  DC molecular Hamiltonian, $\hat{H}_{DC}$, which is:
\begin{equation}
\hat{H}_{DC} = 
\sum_i \left[c{\alpha}\cdot{p}_i + \beta mc^2 - \sum_A \frac{Z_A}{|{r}_i - {R}_A|}\right]
+ \sum_{i\neq j} \frac{1}{|{r}_i - {r}_j|}.
\end{equation}
Here, $c$ is the speed of light, ${\alpha}$ and $\beta$ are the Dirac matrices, and ${p}_i$ refers to the momentum of the $i^{th}$ electron. The summation over the electronic coordinates is denoted by $i$, and that over the nuclear coordinates is indicated by $A$. ${r}_i$ is the position vector from the origin to the site of an electron, and ${R}_A$ is the position vector from the origin to the coordinate of a nucleus. $Z_A$ is the atomic number of the $A^{th}$ nucleus. Then, the second-quantized form of the Hamiltonian is given by:
\begin{equation}
    \hat{H} = \sum_{pq} h_{pq} a_p^\dagger a_q + \frac{1}{2} \sum_{pqrs} g_{pqrs} a_p^\dagger a_q^\dagger a_s a_r .
\end{equation}
where \( h_{pq} \)  and \( g_{pqrs} \) are the  one-and two-electron integrals, respectively and 
\( a_p^\dagger \text{and } a_q \) are the fermionic creation and annihilation operators, and the indices run over relativistic orbitals.
The releativistic equation for molecules  can be expressed as:
$\hat{H} \ket{\Psi} = E \ket{\Psi}$,
where $\hat{H}$ is the DC molecular Hamiltonian, $\ket{\Psi}$ and $E$  are the corresponding wavefunction and  energy  respectively.
The wavefunction $\ket{\Psi}$ is expanded as a 
linear combination of Slater determinants $\ket{\Psi} = \sum_{i} c_i \ket{\Phi_i}$.
The determinatal states $\ket{\Phi_i}$s include the DF mean-field and  many electron particle-hole states. The $c_i$ s are the expansion coefficients.

The molecular PDMs in this study were first calculated using classical computers  to provide 
benchmarks for the values obtained via the QAE algorithm.  Specifically, DF and relativistic coupled-cluster singles and doubles 
(RCCSD) calculations were performed using the DIRAC22 software package~\cite{DIRAC22}. The DF computations 
were carried out at the self--consistent field (SCF) level, with a perturbative electric field ($\epsilon = 
\pm 10^{-3}$) applied along the $z$-axis to evaluate the field-dependent energies. To enhance computational efficiency, 
molecular symmetry was exploited through use of C$_{2v}$ double group symmetry~\cite{Dyall2007}.


To enable a direct comparison with the QAE algorithm, the FCI Hamiltonian was constructed classically by applying the Slater–Condon rules~\cite{slatercondonrules} and subsequently diagonalized exactly to obtain the FCI/ED energy. This  Hamiltonian was 
then mapped to the D-Wave quantum annealer, where the lowest eigenvalue corresponding to the ground-state energy was 
extracted. The ground-state energies corresponding to both positive and negative perturbation strengths ($\epsilon = \pm 10^{-3}$) were computed to evaluate the field-dependent energy shifts. QA was employed to 
determine the electronic energies in active spaces $(No,Me)$  = $(8o, 3e)$ and $(14o, 7e)$ where, 
$M$ denotes the number of occupied orbitals out of 
$N$ total orbitals. These electronic energies were then combined with the corresponding core energy and nuclear repulsion energy terms 
to obtain the total molecular energy. Using this total energy, the electronic component of the dipole moment was 
evaluated through the finite-field approach as shown by Eq.(\ref{eq: ffm}). 

To obtain the PDM, the nuclear contribution was added based on experimentally 
available bond lengths. The bond lengths used in the calculations were 1.361\,\AA\ for BeF, 1.75\,\AA\ for MgF, 
1.967\,\AA\ for CaF, 2.075\,\AA\ for SrF, and 2.16\,\AA\ for BaF~\cite{pra2016, Hao2019, langhoff1986}. For all the calculations, uncontracted Gaussian-type orbitals (GTOs)~\cite{Dunning:1989} were used, and the kinetic balance 
condition was implemented to ensure the proper relationship between the large and small components of the orbitals. Each orbital is represented by  n-Slater type functions, each with a different value of exponent~(zeta)~\cite{Dyall2007}. The basis sets for the heavier 
atoms Sr and Ba were chosen from dyall.c2v, dyall.c3v~\cite{BSE}, and dyall.c4v~\cite{Dyall:2009, Dunning:1989} For lighter elements such as Be, Mg, Ca, and F, using exponents
 from the correlation-consistent polarized valence (cc-pV) basis, which were obtained from the EMSL Basis Set Exchange Library~\cite{BSE}.




QA involves preparing the ground-state of an initial Hamiltonian, $\hat{H}_I$, which is typically chosen to be the transverse field Hamiltonian, and then adiabatically transforming it into another Hamiltonian, $\hat{H}_F$, generally chosen to be the Ising Hamiltonian whose ground-state encodes the solution to the problem. The transformation follows:

\begin{equation}
\hat{H}(t) = f(t)\hat{H}_I + g(t)\hat{H}_F; \quad t \in [0, T],
\end{equation}

where the system evolves smoothly from the ground state of $\hat{H}_I$ to that of $\hat{H}_F$ during the annealing time $T$. Initially, $f(0) \gg g(0)$, giving a simple delocalized ground state with spins $s_i = \pm1$. As time progresses, $f$ decreases and $g$ increases, and by $t = T$, $f(T) \ll g(T)$, completing the annealing process.
While the adiabatic model described above is the theoretical foundation of QA, in practice, quantum annealing operates far from the adiabatic limit. In this work, we apply the QAE algorithm to the electronic structure problem. The QAE algorithm solves the 
eigenvalue equation $\hat{H} \ket{\Psi} = E \ket{\Psi}$ by transforming it into an energy minimization 
problem~\cite{Teplukhin_2020, Teplukhin2021}. The energy functional considered is the expectation value of the Hamiltonian $\hat{H}$ with 
respect to an unknown state $\ket{\Psi}$. To avoid trivial solutions, the normalization constraint $\braket{\Psi}{\Psi} - 1 = 0$ is enforced by including a Lagrange multiplier $\lambda$ into the 
energy functional. In the modified formulation, after dropping irrelevant constants the energy functional will be
$\xi = \bra{\Psi} \hat{H} \ket{\Psi} - \lambda \braket{\Psi}{\Psi}$.
Assuming the wavefunction is expressed as $\ket{\Psi} = \sum_{i=1}^{N} c_i \ket{\Phi_i}$, where $\{\ket{\Phi_i}\}$ stands for a collection of $N$ predefined basis functions (determinatal states), the 
QAE algorithm seeks to find the coefficients $\{c_i\}$, with each $c_i$ lying in the interval $[-1, 1]$, for the energy functional:

\begin{equation}\label{eq: 5}
\xi(\vec{c}, \lambda) = \sum_{i,j=1}^{N} c_i c_j (H_{ij} - \lambda \delta_{ij}) .
\end{equation}

In this context, $H_{ij} \equiv \bra{\Phi_i} \hat{H} \ket{\Phi_j}$ and $\delta_{ij}$ denotes the Kronecker 
delta function. To translate the problem for quantum annealing, we implemented  fixed-point encoding~\cite{Teplukhin2019} 
strategy wherein each coefficient $c_i$ is expressed using $K$ qubits $q_\alpha^i \in \{0, 1\}$:

\begin{equation}\label{eq: 6}
c_i = \sum_{\alpha=1}^{K-1} 2^{\alpha - K} q_\alpha^i - q_K^i.
\end{equation}

This converts the problem from a continuous optimization over $\{c_i\}$ to a discrete optimization over the binary variables $\{q^i_\alpha\}$. Here, the first $K-1$ qubits encode the fractional part, the last qubit $q_K^i$
encodes the sign and thus the each $c_i$ is repesented by $K$ number of binary variables $q^i_\alpha$~\cite{Xia2018, Streif2019, Teplukhin2019}.
The energy functional becomes:

\begin{equation}
\begin{split}
\xi(\vec{q}, \lambda) = \sum_{i,j=1}^{N} \sum_{\alpha,\beta=1}^{K-1} 
&\left( 2^{\alpha + \beta - 2K} q_\alpha^i q_\beta^j - 2^{\alpha + 1 - K} q_\alpha^i + q_K^i q_K^j \right) \\
&\times \left( H_{ij} - \lambda \delta_{ij} \right).
\end{split}
\end{equation}

The proposed mapping approach necessitates an expansion of the Quadratic Unconstrained Binary 
Optimization (QUBO) formulation~\cite{glover2019} from dimensions $(N \times N)$ to $(NK \times NK)$. 
In view of restricted connectivity limitations inherent 
to D-Wave's quantum annealing architecture, the optimization of $\xi(\vec{q}, \lambda)$ is 
implemented through a pipelined framework consisting of sequential $M$ smaller, more tractable sub-QUBO optimizations, where 
each sub-problem of energy functional $\xi_m(\vec{q}^{(m)},\lambda)$ operates on an independent subset of variables.
Each resulting sub-problem is then optimized independently to find a local solution, denoted as $\vec{q}^{(m)}$ for $m \in \{1, \dots, M\}$. The optimization of these sub-QUBOs can be performed using a variety of solvers, including quantum annealers or classical algorithms such as Simulated Annealing~(SA). Although the decomposition methodology permits parallel execution on multiple quantum annealers, in this work the sub-QUBOs are optimized sequentially due to limited processor access, and thus the overall computational demand remains unchanged.
In our methodology, we first sort the coefficients in descending order on the basis of their 
values obtained classically from perturbation theory. These ordered coefficients are then partitioned into systematic 
groupings that form the basis for sub-QUBOs construction to reduce 
the problem size while preserving the essential features of the solution space. These sub-QUBOs are then independently 
submitted to the annealer to obtain the corresponding solution~\cite{vikrant2024}.

The embedding process maps the each  sub-QUBO onto the D-Wave hardware's native topology~\cite{Pelofske_2024, Boothby2019} by forming 
chains of physical qubits to represent single logical variables. We used the default chain coupling 
strength  during annealing ~\cite{DWaveChainStrength:2022}.
During the annealing process, we first map the subproblem onto the Pegasus graph structure of the D-Wave Advantage quantum processing unit~(QPU). For each execution, we acquire a fixed ensemble of 1000 
samples(anneals or number of reads), where each sample represents a specific configuration of the coefficients in the 
qubit state space.

After obtaining the binary solutions from each independent sub-QUBO problem from the quantum annealer, we aggregate those
solutions and proceed to a post-processing stage. This stage involves applying a steepest descent 
optimization at every value of $\lambda$ within the defined search interval $[\lambda_{\text{min}}, 
\lambda_{\text{max}}]$. For each value of $\lambda$ the annealer returns binary values of ${\vec{q}^{(m)}_{\text{opt}}}$ for each sub-QUBO.  In our case, we have set $K = 10$ to encode each parameter with sufficient resolution. The 
entire implementation utilizes the tools available in D-Wave’s Ocean SDK~\cite{DwaveOceanSDK}, and 
specifically employs the D-Wave Advantage  quantum processing units(QPU)
for execution.
Then, the global solution~$(\vec{q}_{\text{raw}})$ of size $NK$ variables is constructed by combining partial solutions~$({\vec{q}^{(m)}_{\text{opt}}})$. We employ steepest descent minimization within the energy landscape 
$\xi(\vec{q}, \lambda)$, where at each iteration the algorithm selects the 
direction of maximal energy reduction. 
This refinement stage compensates for decomposition approximations by explicitly incorporating all inter-connected  couplings from the original Hamiltonian. The converged solution $\vec{q}_{\text{final}}$ attains accuracy comparable to classical methods.

\begin{algorithm}[H]
\caption{QA for Energy Minimization}
\label{alg:quantum_annealing_d-wave}
\begin{algorithmic}[1]
\Require Full Hamiltonian matrix $H$ of size $N \times N$.
\Require Initial guess for the parameter $\lambda \in [\lambda_{\text{min}}, \lambda_{\text{max}}]$.
\Ensure Optimal solution vector $\vec{c}_{\text{final}}$ with minimal energy.

\Statex \textbf{Phase 1: Sub-QUBO Formulation}
\State Partition the Hamiltonian $H$ into $M$ sub-QUBO problems, denoted as $\xi_m(\vec{q}^{(m)}, \lambda)$ for $m \in \{1, \dots, M\}$.
\Statex \textbf{Phase 2: Quantum Annealing Iteration}
\For{$\lambda$ from $\lambda_{\text{min}}$ to $\lambda_{\text{max}}$} \Comment{Iterate over the range of $\lambda$}
    \For{each sub-QUBO $\xi_m(\vec{q}^{(m)}, \lambda)$}
        \State Encode the classical problem into the sub-QUBO: $\xi_m(\vec{q}^{(m)},\lambda) \leftarrow \xi_m(\vec{c}^{(m)},\lambda)$.
        \State Perform quantum annealing on $\xi_m$: $q_{\text{raw}}^{(m)} \leftarrow \min[\xi_m(\vec{q},\lambda)]$ \Comment{Find the optimal binary configuration for the $i$-th sub-QUBO.}
    \EndFor
    \State Combine the optimal solutions from all sub-QUBOs: $\vec{q}_{\text{raw}} \leftarrow \bigcup_{m=1}^M q_{\text{opt}}^{(m)}$.
    
    \Statex \textbf{Phase 3: Post-processing and Energy Evaluation}
    \State Apply a steepest descent method for refinement: $\vec{q}_{\text{final}} \leftarrow \text{SteepestDescent}(\vec{q}_{\text{raw}})$.
    \State Reverse the encoding to obtain classical coefficients: $\vec{c}_{\text{final}} \leftarrow \text{ReverseEncoding}(\vec{q}_{\text{final}})$.
    \State Compute the total energy $E_\lambda$ using the obtained coefficients: $E_\lambda \leftarrow \text{ComputeEnergy}(\vec{c}_{\text{final}})$.
\EndFor

\Statex \textbf{Final Solution Selection}
\State Identify the best solution: $\vec{c}_{\text{final}} \leftarrow \underset{\lambda \in [\lambda_{\text{min}}, \lambda_{\text{max}}]}{\operatorname{argmin}}\{E_\lambda\}$. \Comment{Select the coefficient vector corresponding to the minimum energy found.}
\State \textbf{Return} $\vec{c}_{\text{final}}$

\end{algorithmic}
\end{algorithm}

\input{flowchart.tex}


In Tables~\ref{tab:result_8o_3e} and~\ref{tab:result_14o_7e}, we present our computed PDM values using the QAE and 
benchmark them against conventional methods such as RCCSD, FCI, and SA results~\cite{Kirkpatrick1983}, all within two 
different active spaces but employing different basis sets. The QAE 
computations were performed on both classical (SA) platforms~\cite{Kirkpatrick1983} and quantum annealing hardware (D-Wave Advantage). These simulations explicitly took into account qubit connectivity constraints. The results indicate 
that QAE can accurately predict both total energies (for both perturbation $\pm \epsilon$)~and PDM values, with 
deviations within acceptable limits. Specifically, the hardware-executed QAE results were found to be consistent with 
their SA and FCI counterparts as given in Table~\ref{tab:errors}.


For the QAE calculations, we selected energy ranges $[\lambda_{\text{min}}, 
\lambda_{\text{max}}]$ slightly below and above the DF
energy, corresponding to the active spaces $(8o,3e)$ and $(14o,7e)$ used in this study. After generating the full 
configuration interaction  determinant basis, we employed a perturbative selection procedure to retain only the most 
relevant configurations. This step mitigates the exponential growth in the number of Slater determinants with 
increasing numbers of virtual orbitals. The Hamiltonian matrix was then constructed 
within the reduced determinant space and used for subsequent QAE calculations.
For each specified energy range $[\lambda_{\text{min}}, 
\lambda_{\text{max}}]$, we performed three independent QAE runs which we term as repeats 
and selected the lowest energy~\cite{zade2025, Teplukhin2019}. Additionally, sub-QUBOs were restricted to a maximum of 60 binary variables for practical implementation on the hardware in this calculation.

We now discuss possible sources of error in our calculation~\cite{DWaveChainStrength:2022, Pearson2019}. These include 
hardware-induced errors, chain-breaking artifacts, and approximations made during sub-QUBOs generation~\cite{amin2023quantumerrormitigationquantum}. In particular, 
during sub-QUBOs construction, certain quadratic interaction terms may be  neglected, potentially impacting the final 
energy estimates. We also performed the same computations using simulated annealing on the classical processor with 
the same sub-QUBO settings and observed a negligible error, suggesting strong agreement between quantum and classical 
annealing techniques.
Assuming that SA and ideal QA yield similar outcomes, we conclude that hardware noise 
likely has a minimal impact on the accuracy of PDMs. This stability may 
result from the use of a large number of anneals (e.g., 1000), which effectively average out random fluctuations and 
reduce noise effects.

Furthermore, we conducted simulations assuming all-to-all connectivity and using the same sub-QUBO decomposition as in 
the quantum hardware setup. This residual discrepancy is primarily attributed to limitations in our QUBO encoding 
scheme—specifically, the choice of $K = 10$ in the fixed-point representation of the expansion coefficients, yielding a 
precision of $\frac{1}{2^{10}}$ in energy. Increasing $K$ would improve numerical precision but also significantly 
enlarge the QUBO, rendering it less feasible for hardware execution at present.
The most prominent deviation was observed between simulations assuming all-to-all connectivity~\cite{Boothby2021} and 
those constrained by the native connectivity of the D-Wave hardware, indicating that chain formation and embedding 
limitations are key sources of inaccuracy~\cite{DWaveChainStrength:2022}.

Finally, we note that our method does not seems to suffer from barren plateau and local minima, which are common in approaches like 
VQE~\cite{barrenplateau2025, Peruzzo2014}. In QAE, the energy functional is optimized to obtain the qubit binary variables, from which the 
expansion coefficients are subsequently determined, as shown in Eq.~(\ref{eq: 6}). Since this approach does not involve 
gate-based operations, it is free from gate errors that typically affect the performance of variational algorithms like 
VQE. Moreover, for the SrF molecule in the $(14o, 7e)$ active space, our method achieved an accuracy of about $98.6\%$,
while comparable accuracy in VQE would require extensive circuit optimization and error mitigation techniques
as exemplified by a 6-qubit VQE computation on IonQ hardware, where an accuracy of $93.1\%$ was achieved~\cite{Palak2024}.
The accurate computation of properties such as the PDM further demonstrates the robustness and reliability of QAE compared 
to VQE.

\begin{table}[h]
\caption{Comparison of the PDM values, in Debye, for the $(8o, 3e)$  active space. Results from the QAE are benchmarked against classical RCCSD, FCI, and SA methods for the molecules and basis sets studied.}
\label{tab:result_8o_3e}
\vspace{3pt}
\begin{tabular}{lc|cccc}
\hline
Molecule & Basis & \ \ RCCSD \ \  & \ \ FCI \ \  & \ \ SA \ \  & \ \ QAE \ \  \\
\hline
BeF & cc-pVDZ & 1.338 & 1.342 & 1.351 & 1.354 \\
    & cc-pVTZ & 1.242 & 1.245 & 1.239 & 1.267 \\
    & cc-pVQZ & 1.270 & 1.271 & 1.273 & 1.267 \\
\hline
MgF & cc-pVDZ & 3.188 & 3.185 & 3.189 & 3.178 \\
    & cc-pVTZ & 3.073 & 3.045 & 3.114 & 3.074 \\
    & cc-pVQZ & 3.099 & 3.099 & 3.098 & 3.108 \\
\hline
CaF & cc-pVDZ & 2.685 & 2.688 & 2.686 & 2.684 \\
    & cc-pVTZ & 2.687 & 2.685 & 2.686 & 2.688 \\
    & cc-pVQZ & 2.684 & 2.683 & 2.680 & 2.681 \\
\hline
SrF & dyall.v2z & 2.753 & 2.753 & 2.752 & 2.752 \\
    & dyall.v3z & 2.723 & 2.723 & 2.721 & 2.719 \\
    & dyall.v4z & 2.759 & 2.759 & 2.763 & 2.764 \\
\hline
BaF & dyall.v2z & 2.323 & 2.323 & 2.323 & 2.322 \\
    & dyall.v3z & 2.222 & 2.222 & 2.222 & 2.222 \\
    & dyall.v4z & 2.297 & 2.297 & 2.296 & 2.293 \\
\hline
\end{tabular}
\end{table}

\begin{table}[h]
\caption{Comparison of the PDM values, in Debye, for the $(14o, 7e)$ active space. Results from the QAE are benchmarked against classical RCCSD, FCI, and SA methods for the molecules and basis sets studied.}
\label{tab:result_14o_7e}
\vspace{3pt}
\begin{tabular}{lc|cccc}
\hline
Molecule & Basis & \ \ RCCSD \ \  & \ \ FCI \ \  & \ \ SA \ \  & \ \ QAE \ \  \\
\hline
BeF & cc-pVDZ & 1.318 & 0.421 & 0.471 & 0.635 \\
    & cc-pVTZ & 1.318 & 0.418 & 0.448 & 0.914 \\
    & cc-pVQZ & 1.268 & 1.268 & 1.275 & 1.267 \\
\hline
MgF & cc-pVDZ & 3.175 & 3.222 & 3.160 & 3.218 \\
    & cc-pVTZ & 3.072 & 3.072 & 3.084 & 3.078 \\
    & cc-pVQZ & 3.099 & 3.105 & 3.109 & 3.103 \\
\hline
CaF & cc-pVDZ & 2.700 & 2.686 & 2.712 & 2.654 \\
    & cc-pVTZ & 2.694 & 2.688 & 2.685 & 2.695 \\
    & cc-pVQZ & 2.690 & 2.684 & 2.681 & 2.680 \\
\hline
SrF & dyall.v2z & 2.764 & 2.759 & 2.750 & 2.755 \\
    & dyall.v3z & 2.734 & 2.730 & 2.727 & 2.770 \\
    & dyall.v4z & 2.764 & 2.762 & 2.762 & 2.724 \\
\hline
BaF & dyall.v2z & 2.324 & 2.322 & 2.315 & 2.322 \\
    & dyall.v3z & 2.225 & 2.223 & 2.224 & 2.219 \\
    & dyall.v4z & 2.299 & 2.298 & 2.296 & 2.298 \\
\hline
\end{tabular}
\end{table}

\begin{table}[h]
\caption{Percentage error for the PDM values calculated via the QAE and SA. The error is calculated relative to the benchmark FCI results presented in Tables II and III for the $(8o, 3e)$ and $(14o, 7e)$ active spaces, respectively.}
\label{tab:errors}
\small
\begin{tabular}{lc|cc|cc}
\hline
Molecule & Basis & \multicolumn{2}{c}{(8o,3e)} & \multicolumn{2}{c}{(14o,7e)} \\
         &       & $\mathrm{Err_{QAE}}$ & $\mathrm{Err_{SA}}$ & $\mathrm{Err_{QAE}}$ & $\mathrm{Err_{SA}}$ \\
\hline
BeF & cc-pVDZ & 0.894 & 0.693 & 50.837 & 11.738 \\
    & cc-pVTZ & 1.815 & 0.423 & 118.834 & 7.177 \\
    & cc-pVQZ & 0.279 & 0.153 & 0.116 & 0.565 \\
\hline
MgF & cc-pVDZ & 0.213 & 0.122 & 0.128 & 1.927 \\
    & cc-pVTZ & 0.940 & 2.241 & 0.180 & 0.381 \\
    & cc-pVQZ & 0.312 & 0.007 & 0.089 & 0.116 \\
\hline
CaF & cc-pVDZ & 0.153 & 0.090 & 1.169 & 0.998 \\
    & cc-pVTZ & 0.122 & 0.032 & 0.264 & 0.128 \\
    & cc-pVQZ & 0.086 & 0.131 & 0.148 & 0.099 \\
\hline
SrF & dyall.v2z & 0.037 & 0.031 & 0.154 & 0.341 \\
    & dyall.v3z & 0.164 & 0.091 & 1.469 & 0.085 \\
    & dyall.v4z & 0.180 & 0.164 & 1.354 & 0.006 \\
\hline
BaF & dyall.v2z & 0.024 & 0.005 & 0.017 & 0.340 \\
    & dyall.v3z & 0.011 & 0.006 & 0.221 & 0.027 \\
    & dyall.v4z & 0.195 & 0.052 & 0.004 & 0.118 \\
\hline
\end{tabular}
\end{table}

In summary, we have demonstrated the efficacy of QAE algorithm in computing the  PDMs of molecules. Unlike  the traditional 
applications of  QAE, which focus on finding the ground-state energy, our approach integrates perturbative techniques 
and the FFM to enable the calculation of molecular properties. This novel framework holds promise for 
a broad range of many-body physics applications. Key contributions of our work include the integration of the finite-field approach into the QAE workflow and the division of the full QUBO into smaller, independent sub-QUBOs by selecting important expansion coefficients. These sub-QUBOs can be annealed in parallel and subsequently merged for post-processing to recover the final 
configuration.

The calculated PDM values for  MgF, CaF, SrF, and BaF exhibit excellent agreement with classical benchmark methods, 
including FCI and RCCSD results, with deviations  of 
average  below 1\%. The results are  in agreement with those obtained from SA. 
An important feature of relativistic coupled-cluster method is that it includes physical effects to all orders of perturbation in the residual Coulomb interaction for all  n particle-n hole excitations.
Tables \textcolor{blue}{I} and \textcolor{blue}{II} of supplementary material indicate that absolute errors relative to FCI are 
consistently on the order of $10^{-5}\%$ or less. 
The analysis also shows a few anomalies in the PDM calculation. For the BeF molecule in the larger 
(14o, 7e) active space, 
QAE performs less accurately with some basis sets, possibly due to how the molecule’s electronic 
structure interacts with the sub-QUBO partitioning. Aside from these few cases, the method works 
well across different molecules and basis sets.

Importantly, this work represents the first computation of a molecular property other than energy using a quantum annealer. Our findings indicate that QAE, when combined with finite-field methods and perturbative many-electron basis selection, could serve as a reliable tool for evaluating electronic properties of molecules.
Scaling up QAE to larger active spaces presents challenges due to increased encoding 
    complexity and subtle convergence issues~\cite{Booth2017, DWaveHybrid}. However, recent advances in quantum-classical workflows can treat an individual orbital as a qubit as in,  variational quantum 
annealing~(varQA) algorithm~\cite{yip2025variationalquantumannealingquantum}.
Future advances in quantum annealing hardware, such as new D-Wave Advantage2 QPU 
with Zephyr topology~\cite{Boothby2021} and non-linear quadratic 
solver~\cite{osaba2024dwavesnonlinearprogramhybridsolver}, may further reduce embedding overheads and 
improve solution accuracy. These developments are expected to enhance both the QUBO encoding and 
annealing stages of QAE. Quantum annealers may soon play a significant role 
in atomic and molecular computations, enabling studies beyond the capabilities of classical 
methods with continued improvements in hardware and algorithm 
design—including better preprocessing and error mitigation 
techniques~\cite{amin2023quantumerrormitigationquantum, 
djidjev2024replicationbasedquantumannealingerror}.


\begin{acknowledgments}
\textit{Acknowledgment: }PPS acknowledges Vikrant Kumar and Aashna Anil Zade for useful discussions on the D-Wave quantum annealer, and Palak Chawla for discussions related to the DIRAC22 package.
\end{acknowledgments}

\bibliographystyle{apsrev4-2}
\bibliography{ref2.bib}

\input{supp}

\end{document}

%% file: flowchart.tex
\begin{figure}[t]
\centering
\begin{tikzpicture}[
    every node/.style={font=\scriptsize, align=center},
    process/.style={
        rectangle, rounded corners,
        draw=black!70, fill=blue!6,
        text width=2.8cm, minimum height=0.9cm,
        inner sep=3pt
    },
    arrow/.style={
        -{Latex[length=1.8mm,width=1mm]},
        thick, draw=cyan!60!black
    }
]

\node[process] (energyfunctional) at (90:2.4cm)
    {\textbf{Construct sub-QUBOs}};
\node[process] (qubo) at (18:2.4cm)
    {\textbf{sub-QUBO energy functional}\\[2pt] $\xi_m(\vec{q}^{(m)}, \lambda)$};
\node[process] (optimization) at (-45:2.6cm)
    {\textbf{Optimization}\\[2pt] Embedding \& annealing on QPU + postprocessing};
\node[process] (post) at (-135:2.6cm)
    {\textbf{Wavefunction construction}\\[2pt] Build $|\Psi\rangle$ from qubit configuration};
\node[process] (energy) at (162:2.4cm)
    {\textbf{Energy evaluation}\\[2pt] $\langle \Psi | \hat{H} | \Psi \rangle$};

\draw[arrow, bend left=12] (energyfunctional) to (qubo);
\draw[arrow, bend left=12] (qubo) to (optimization);
\draw[arrow, bend left=12] (optimization) to (post);
\draw[arrow, bend left=12] (post) to (energy);
\draw[arrow, bend left=12] (energy) to (energyfunctional);

\draw[arrow, bend right=15]
    ([xshift=-0.45cm, yshift=0.5cm]energyfunctional.west)
    to node[midway, above, font=\bfseries\tiny, black!90]
    {$\lambda \in [\lambda_{\min}, \lambda_{\max}]$}
    ([xshift=-0.25cm, yshift=0.3cm]energyfunctional.west);

\end{tikzpicture}

\caption{Workflow of the QAE algorithm, where an energy functional with Lagrange multiplier $\lambda$ is mapped to a sub-QUBO, optimized on the quantum annealer, and refined via post-processing across $\lambda$ values to obtain the minimum ground-state energy.}
\label{fig:flowchart}
\end{figure}
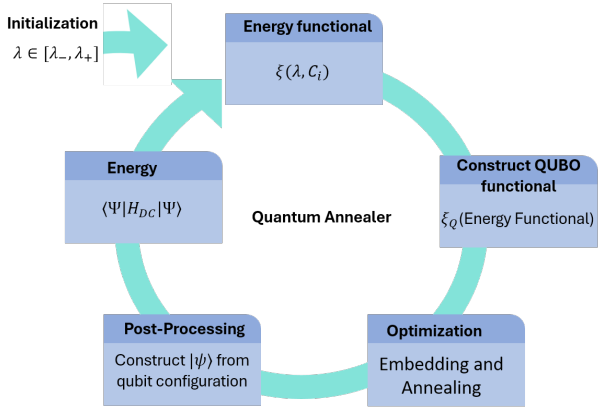

%% file: supp.tex
\onecolumngrid 
\appendix
\section{SUPPLEMENTARY MATERIAL : Ground-State Energy Benchmark Data}

\begin{table}[h]
\centering
\caption{Ground-state energy comparison (in Hartree) for the (8o, 3e) active space under an applied electric field perturbation of $\epsilon=\pm0.001$ a.u. QAE and SA results are compared to the benchmark FCI energies, with absolute errors reported.}
\label{tab:energy_comparison1}
\begin{tabular}{l |l |rr| r r| r r| r r}
\hline
Molecule & Basis & \multicolumn{2}{c}{FCI} & \multicolumn{2}{c}{SA} & \multicolumn{2}{c}{QAE} & \multicolumn{2}{c}{\% Error in QAE } \\
& &$\epsilon=-0.001$&$\epsilon=+0.001$& $\epsilon=-0.001$&$\epsilon=+0.001$& $\epsilon=-0.001$&$\epsilon=+0.001$& $\epsilon=-0.001$&$\epsilon=+0.001$ \\
\hline
BeF & cc-pVDZ & -114.218092 & -114.198572 & -114.218082 & -114.198571 & -114.218075 & -114.198565 & 1.45E-05 & 6.19E-06 \\
    & cc-pVTZ & -114.263250 & -114.243655 & -114.263246 & -114.243646 & -114.274282 & -114.254704 & 9.65E-03 & 9.67E-03 \\
    & cc-pVQZ & -114.274286 & -114.254711 & -114.274283 & -114.254709 & -114.274282 & -114.254704 & 3.60E-06 & 6.04E-06 \\
\hline
MgF & cc-pVDZ & -299.548212 & -299.471350 & -299.548205 & -299.471345 & -299.548208 & -299.471340 & 1.45E-06 & 3.23E-06 \\
    & cc-pVTZ & -299.590767 & -299.513813 & -299.590766 & -299.513812 & -299.590766 & -299.513812 & 4.59E-07 & 3.29E-07 \\
    & cc-pVQZ & -299.602748 & -299.525817 & -299.602744 & -299.525813 & -299.602739 & -299.525816 & 3.04E-06 & 5.05E-07 \\
\hline
CaF & cc-pVDZ & -779.318357 & -779.171788 & -779.318355 & -779.171784 & -779.318355 & -779.171783 & 2.12E-07 & 6.28E-07 \\
    & cc-pVTZ & -779.370339 & -779.223768 & -779.370338 & -779.223768 & -779.370336 & -779.223768 & 3.87E-07 & 5.64E-08 \\
    & cc-pVQZ & -779.382993 & -779.236420 & -779.382990 & -779.236415 & -779.382990 & -779.236416 & 3.31E-07 & 5.64E-07 \\
\hline
SrF & dyall.v2z & -3277.855874 & -3277.560031 & -3277.855871 & -3277.560027 & -3277.855871 & -3277.560027 & 9.57E-08 & 1.20E-07 \\
    & dyall.v3z & -3277.907353 & -3277.611486 & -3277.907352 & -3277.611483 & -3277.907353 & -3277.611483 & 9.60E-09 & 1.17E-07 \\
    & dyall.v4z & -3277.917885 & -3277.622046 & -3277.917880 & -3277.622044 & -3277.917881 & -3277.622046 & 1.22E-07 & 2.99E-09 \\
\hline
BaF & dyall.v2z & -8235.802103 & -8235.346769 & -8235.802100 & -8235.346766 & -8235.802101 & -8235.346766 & 3.28E-08 & 3.81E-08 \\
    & dyall.v3z & -8235.856833 & -8235.401419 & -8235.856831 & -8235.401417 & -8235.856830 & -8235.401416 & 3.51E-08 & 3.27E-08 \\
    & dyall.v4z & -8235.868050 & -8235.412695 & -8235.868049 & -8235.412693 & -8235.868049 & -8235.412690 & 1.79E-08 & 6.08E-08 \\
\hline
\end{tabular}
\end{table}

\begin{table}[h]
\centering
\caption{Ground-state energy comparison (in Hartree) for the (14o, 7e) active space under an applied electric field perturbation of $\epsilon=\pm0.001$ a.u. QAE and SA results are compared to the benchmark FCI energies, with absolute errors reported.}

\label{tab:energy_comparison2}
\begin{tabular}{l |l |rr| r r| r r| r r}
\hline
Molecule & Basis & \multicolumn{2}{c}{FCI} & \multicolumn{2}{c}{SA} & \multicolumn{2}{c}{QAE} & \multicolumn{2}{c}{\% error in QAE} \\
& &$\epsilon=-0.001$&$\epsilon=+0.001$& $\epsilon=-0.001$&$\epsilon=+0.001$& $\epsilon=-0.001$&$\epsilon=+0.001$& $\epsilon=-0.001$&$\epsilon=+0.001$ \\
\hline
BeF & cc-pVDZ & -114.219629 & -114.199386 & -114.219477 & -114.199272 & -114.219306 & -114.199230 & 2.83E-04 & 1.36E-04 \\
    & cc-pVTZ & -114.219614 & -114.199368 & -114.219484 & -114.199261 & -114.234397 & -114.214541 & 1.29E-02 & 1.33E-02 \\
    & cc-pVQZ & -114.274486 & -114.254908 & 114.274458 & -114.254887 & -114.274449 & -114.254870 & 3.25E-05 & 3.36E-05 \\
\hline
MgF & cc-pVDZ & -299.548736 & -299.471903 & -299.548732 & -299.471849 & -299.548577 & -299.471740 & 5.31E-05 & 5.42E-05 \\
    & cc-pVTZ & -299.590879 & -299.513928 & -299.590851 & -299.513909 & -299.590841 & -299.513894 & 1.28E-05 & 1.13E-05 \\
    & cc-pVQZ & -299.602824 & -299.525899 & -299.602798 & -299.525876 & -299.602797 & -299.525870 & 8.91E-06 & 9.63E-06 \\
\hline
CaF & cc-pVDZ & -779.318452 & -779.171882 & -779.318414 & -779.171865 & -779.318432 & -779.171836 & 2.63E-06 & 5.80E-06 \\
    & cc-pVTZ & -779.370364 & -779.223796 & -779.370362 & -779.223791 & -779.370357 & -779.223794 & 9.54E-07 & 2.37E-07 \\
    & cc-pVQZ & -779.383009 & -779.236437 & -779.383007 & -779.236433 & -779.383006 & -779.236431 & 3.43E-07 & 7.45E-07 \\
\hline
SrF & dyall.v2z & -3277.888392 & -3277.592554 & -3277.888370 & -3277.592524 & -3277.888377 & -3277.592535 & 4.60E-07 & 5.62E-07 \\
    & dyall.v3z & -3277.939849 & -3277.643987 & -3277.939839 & -3277.643975 & -3277.939844 & -3277.643978 & 1.44E-07 & 2.76E-07 \\
    & dyall.v4z & -3277.950378 & -3277.654541 & -3277.950373 & -3277.654536 & -3277.950371 & -3277.654541 & 1.97E-07 & 8.27E-11 \\
\hline
BaF & dyall.v2z & -8236.102952 & -8235.647617 & -8236.102945 & -8235.647603 & -8236.102943 & -8235.647608 & 1.12E-07 & 1.15E-07 \\
    & dyall.v3z & -8236.157661 & -8235.702248 & -8236.157655 & -8235.702243 & -8236.157653 & -8235.702236 & 9.86E-08 & 1.46E-07 \\
    & dyall.v4z & -8236.168872 & -8235.713518 & -8236.168870 & -8235.713514 & -8236.168867 & -8235.713513 & 6.17E-08 & 6.08E-08 \\
\hline
\end{tabular}
\end{table}

\begin{figure}[H]
    \centering
    \setlength{\tabcolsep}{2pt} 
    \renewcommand{\arraystretch}{0} 
    \begin{tabular}{cc}
    \includegraphics[width=0.48\linewidth,height=4.8cm]{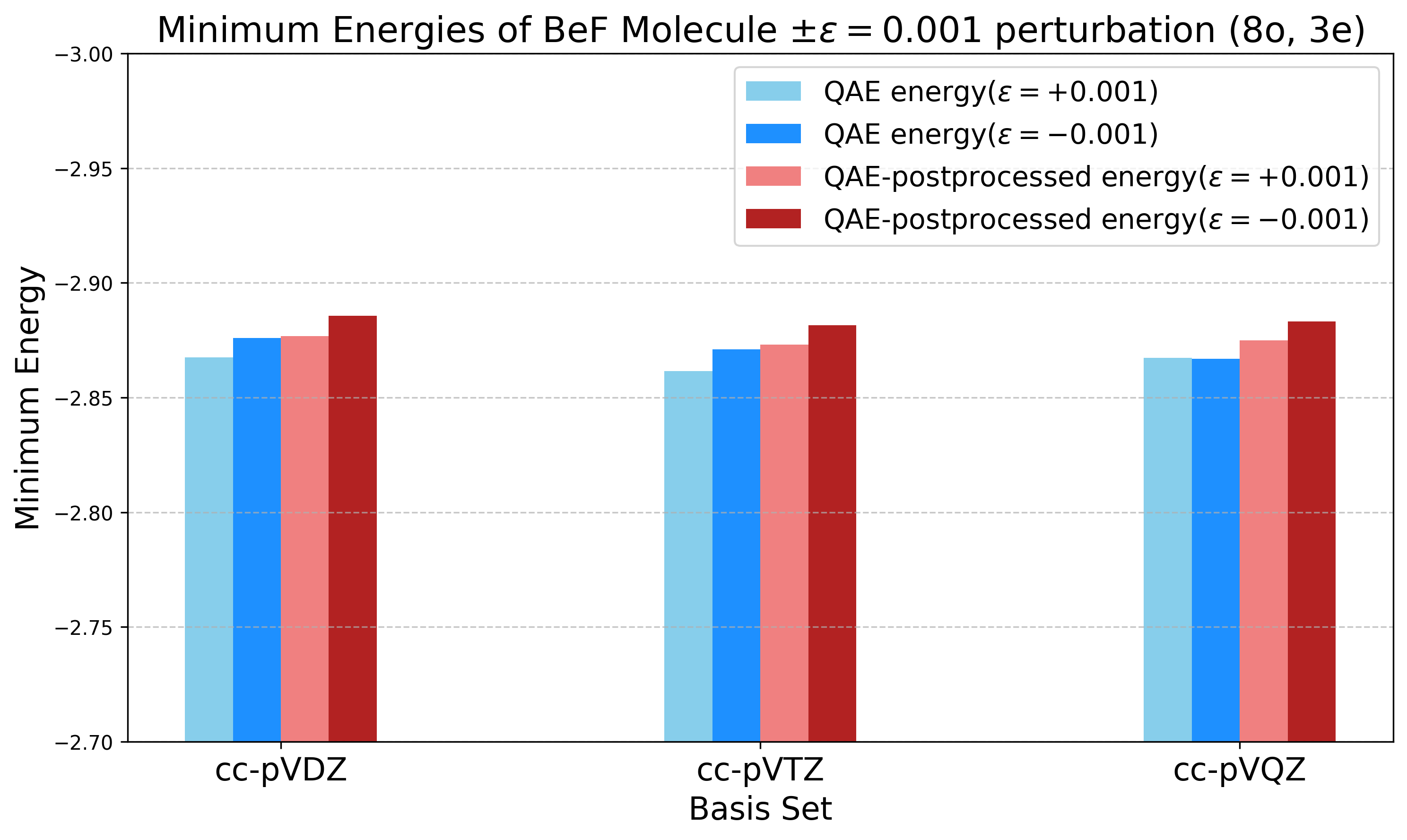} &
    \includegraphics[width=0.48\linewidth,height=4.8cm]{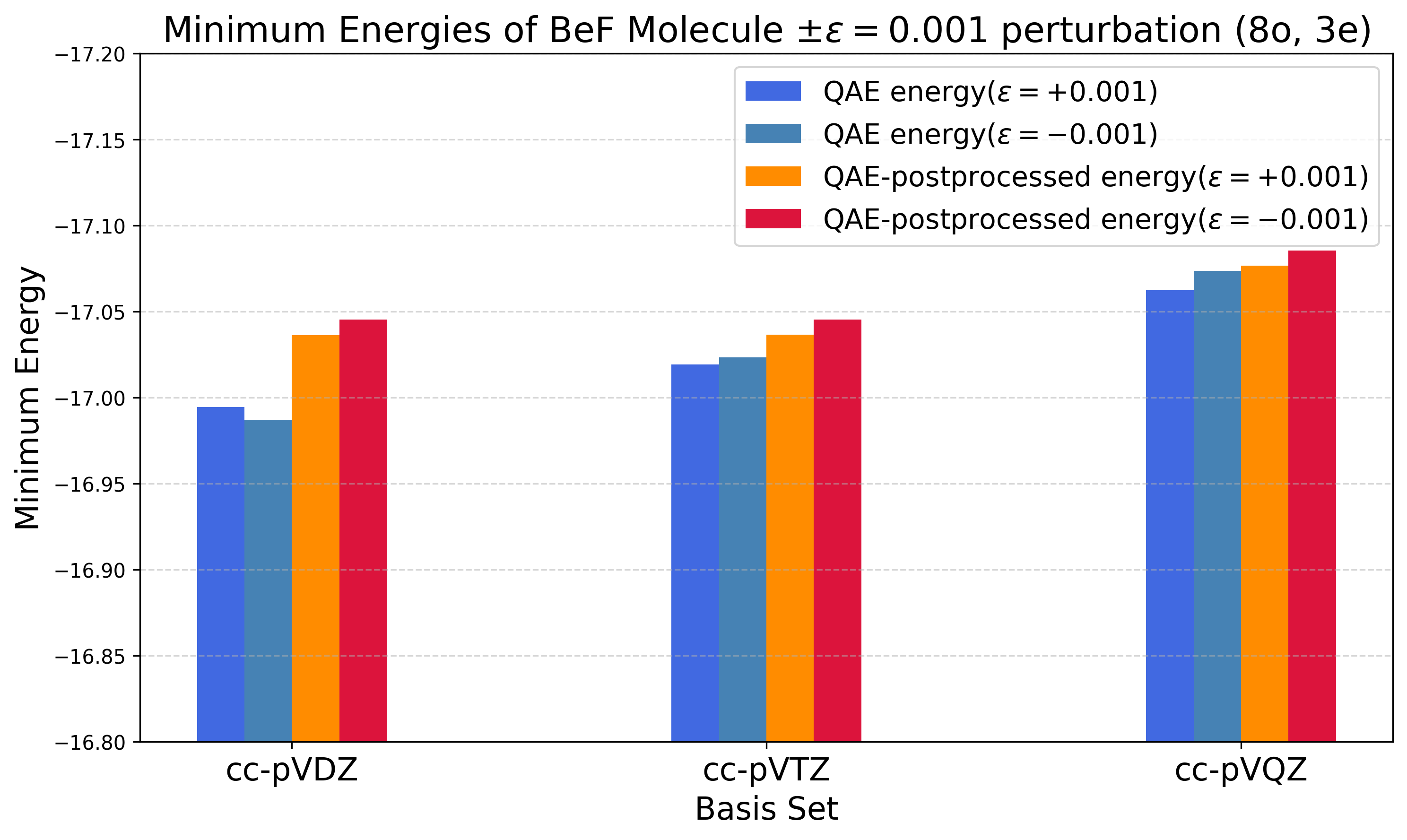} \\
    \scriptsize BeF (8o,3e) & \scriptsize BeF (14o,7e) \\[0.1cm]
    
    


    \end{tabular}
\end{figure}

\begin{figure}[H]
    \centering
    \setlength{\tabcolsep}{2pt} 
    \renewcommand{\arraystretch}{0} 
    \begin{tabular}{cc}
    
    \includegraphics[width=0.48\linewidth,height=4.8cm]{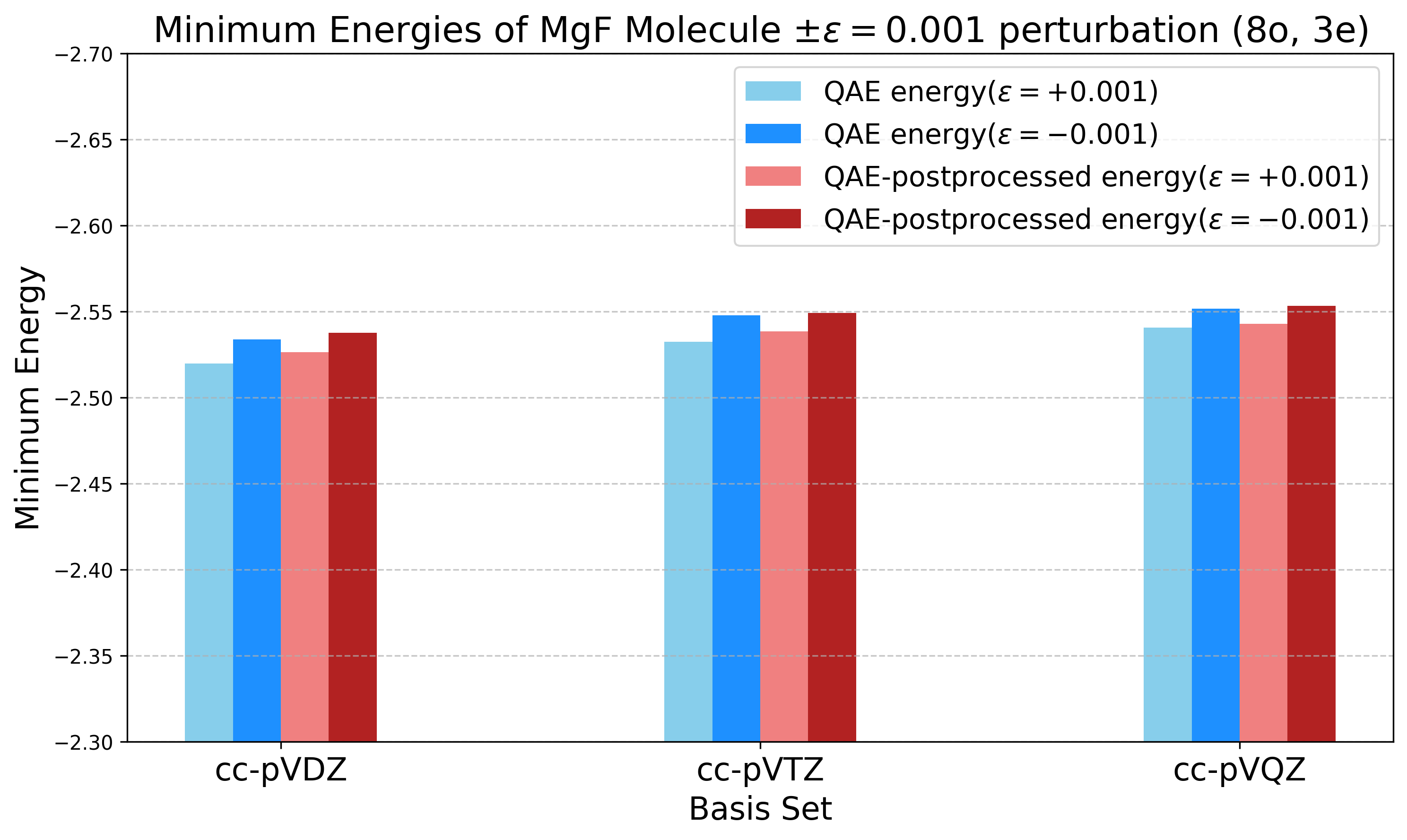} &
    \includegraphics[width=0.48\linewidth,height=4.8cm]{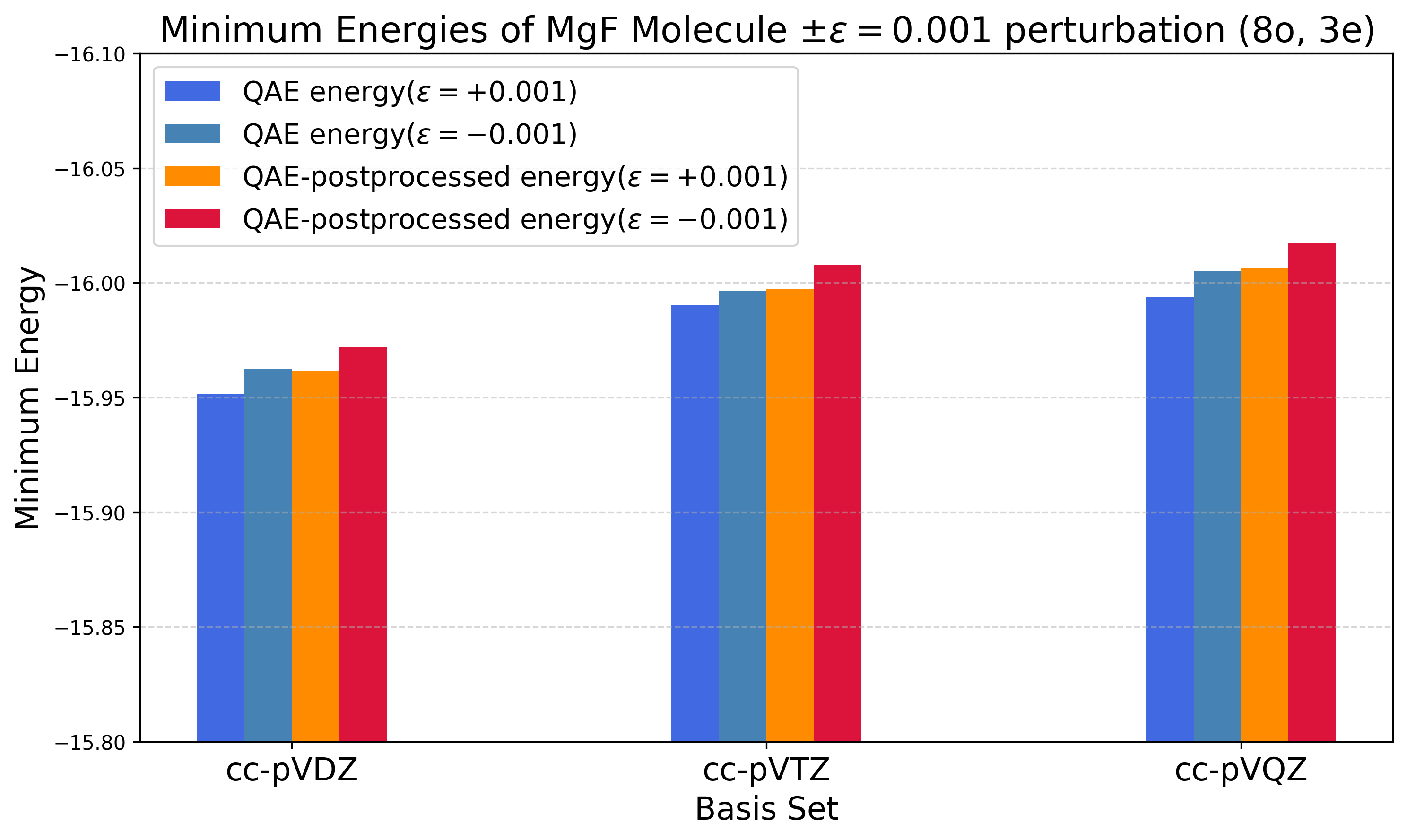} \\
    \scriptsize MgF (8o,3e) & \scriptsize MgF (14o,7e) \\[0.1cm]
    
    \includegraphics[width=0.48\linewidth,height=4.8cm]{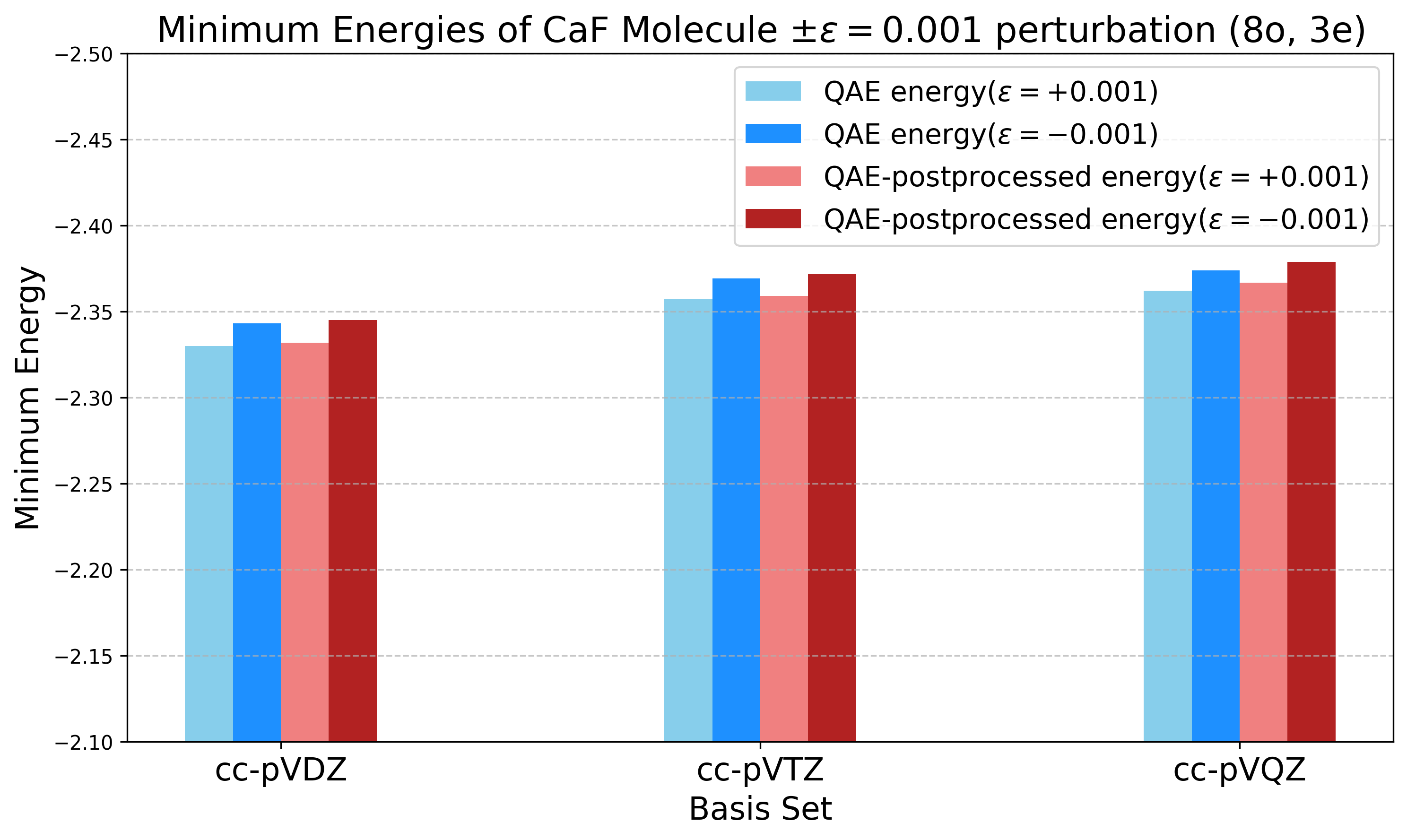} &
    \includegraphics[width=0.48\linewidth,height=4.8cm]{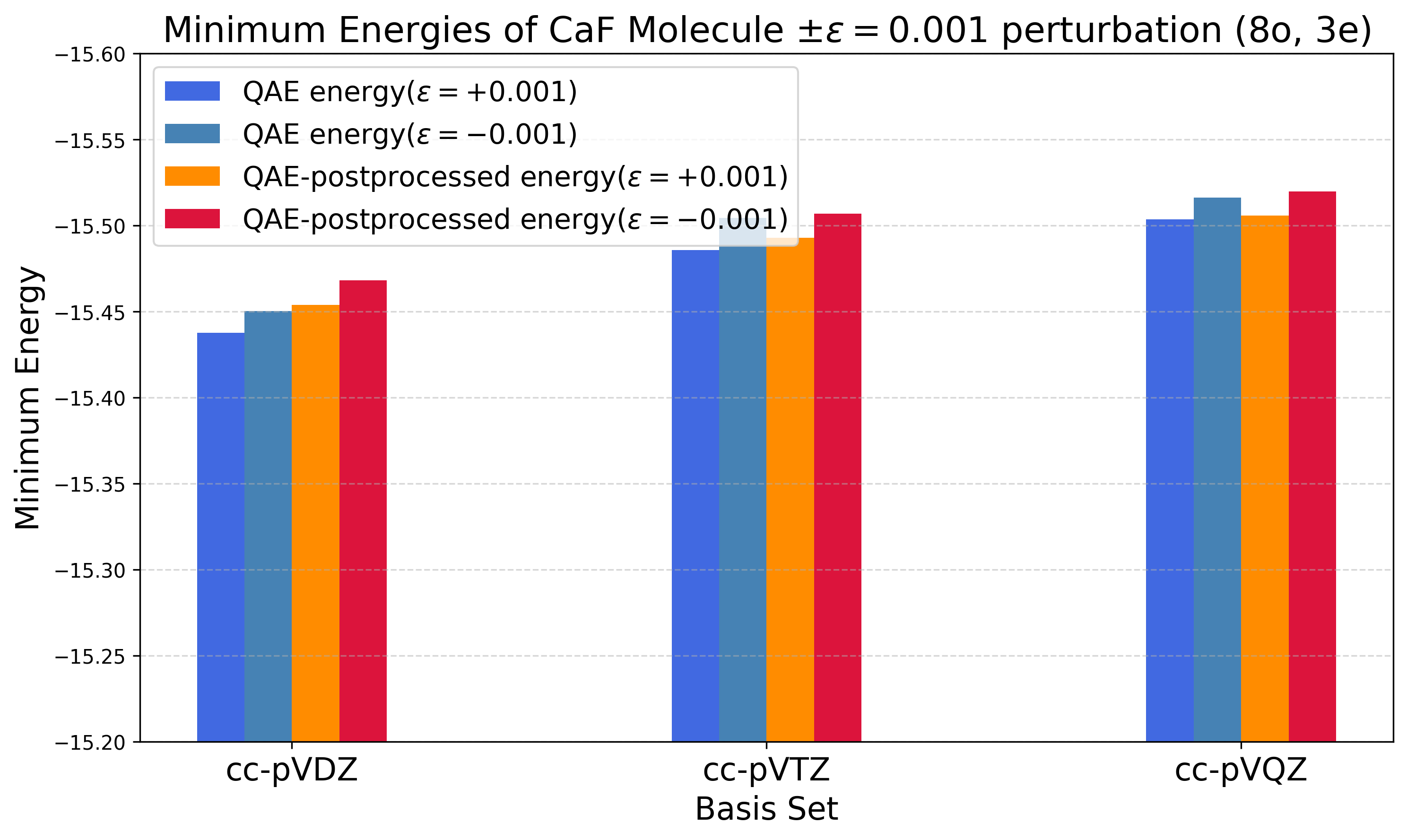} \\
    \scriptsize CaF (8o,3e) & \scriptsize CaF (14o,7e) \\[0.1cm]
    
    \includegraphics[width=0.48\linewidth,height=4.8cm]{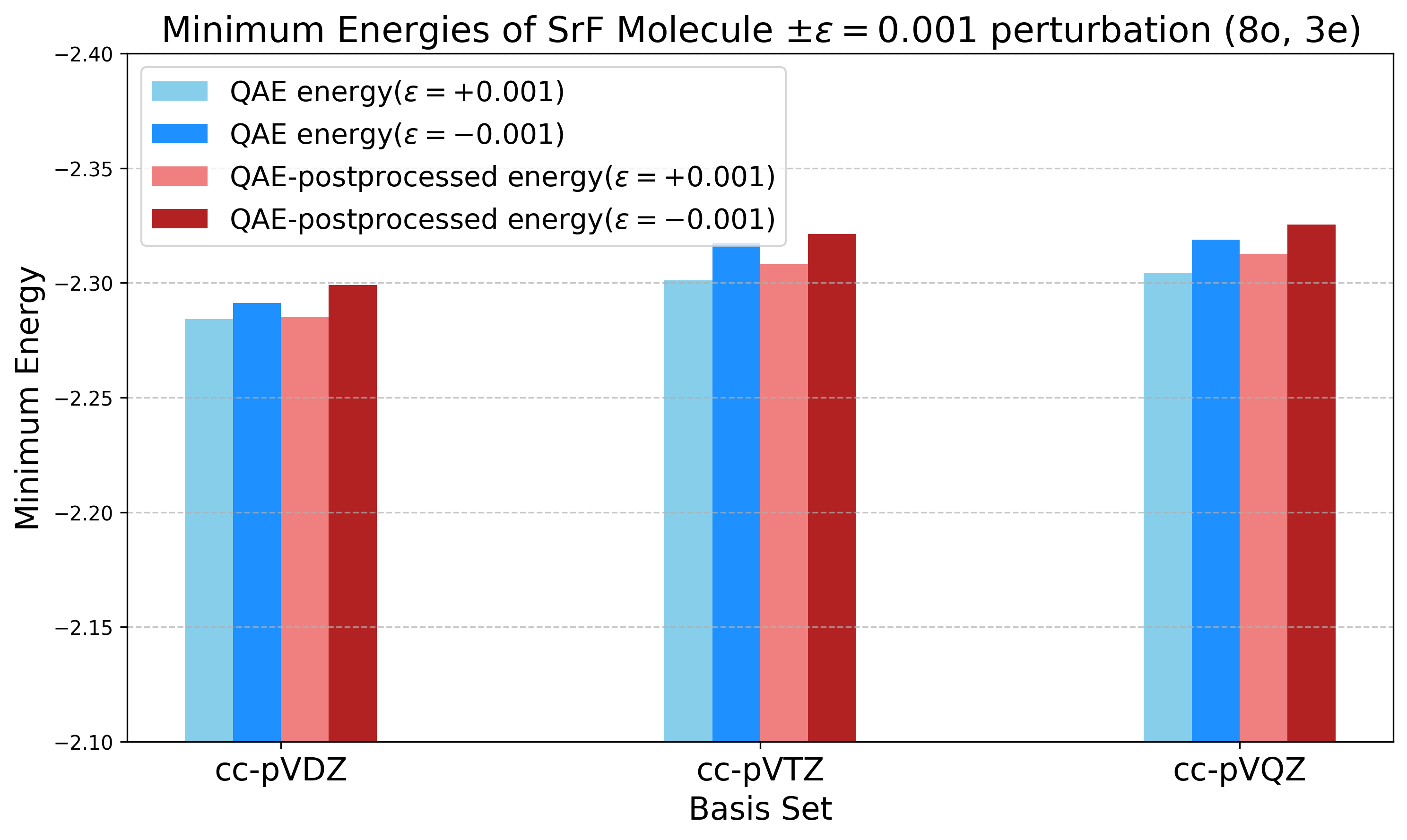} &
    \includegraphics[width=0.48\linewidth,height=4.8cm]{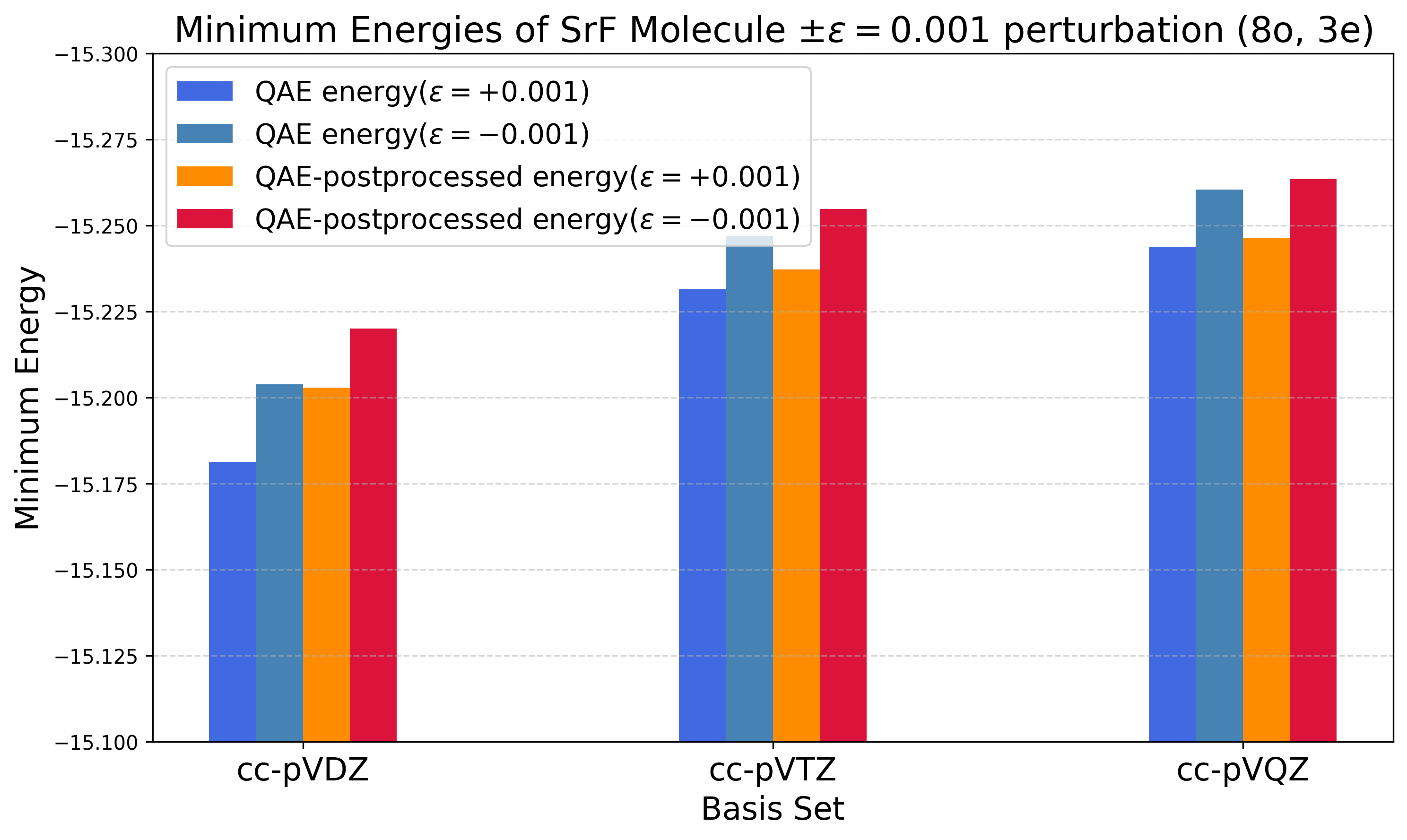} \\
    \scriptsize SrF (8o,3e) & \scriptsize SrF (14o,7e) \\[0.1cm]
    
    \includegraphics[width=0.48\linewidth,height=4.8cm]{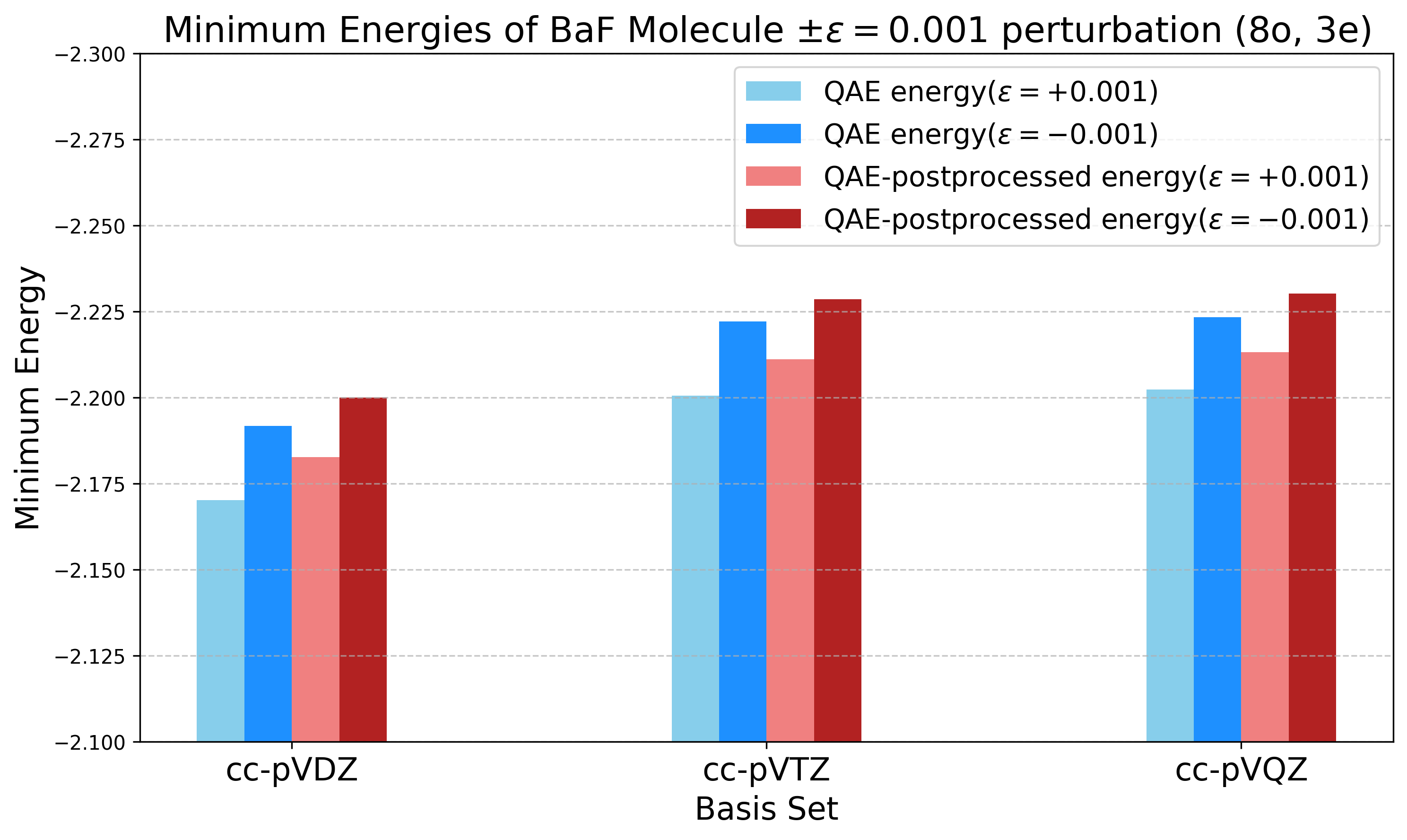} &
    \includegraphics[width=0.48\linewidth,height=4.8cm]{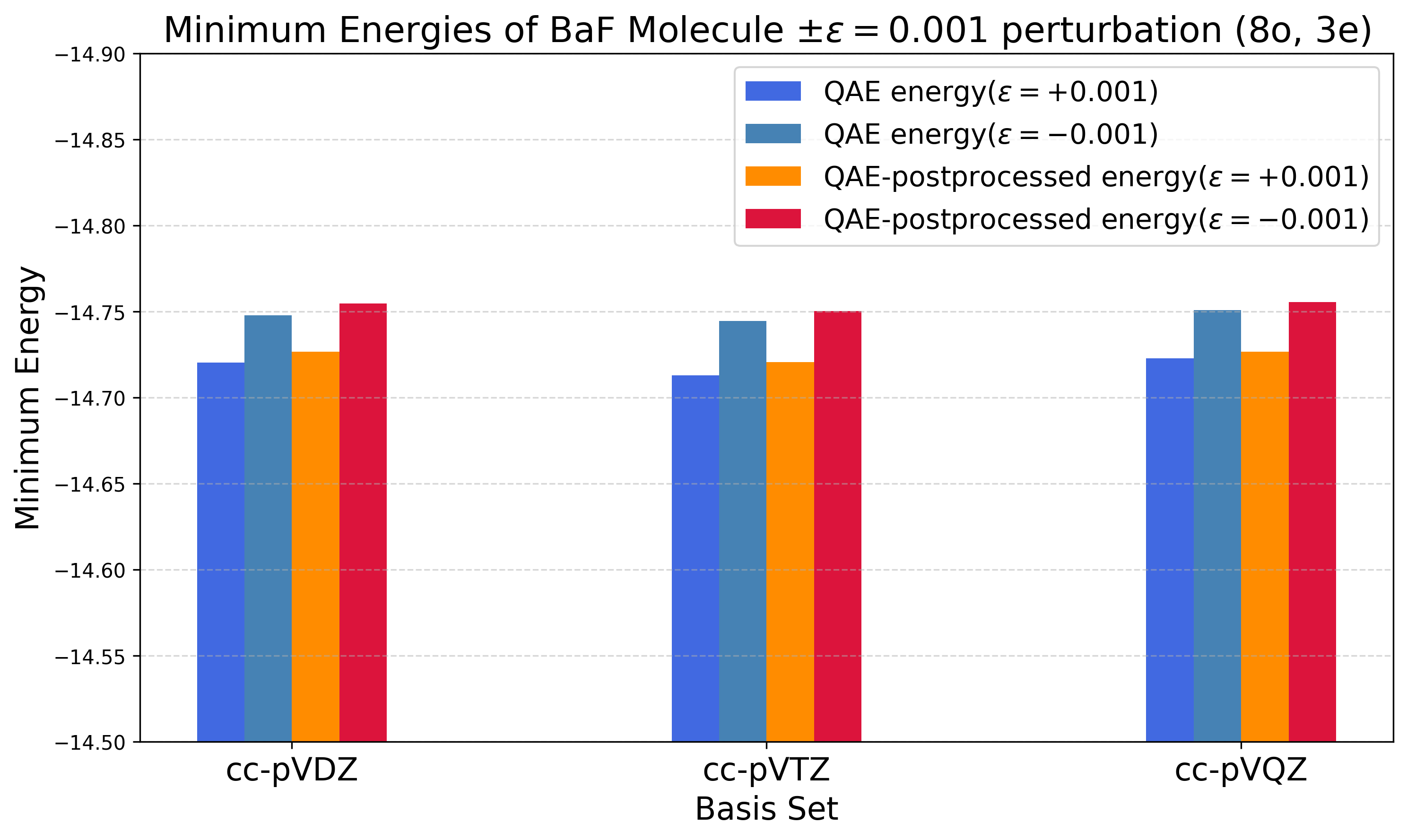} \\
    \scriptsize BaF (8o,3e) & \scriptsize BaF (14o,7e) \\ [0.1cm]
    \end{tabular}
    \caption{Comparison of ground-state energies for molecules under electric field perturbations of $\epsilon=\pm0.001$ a.u. The charts compare the raw output from the annealer (QAE energy), the energy after classical refinement (QAE-postprocessed energy), and the benchmark FCI energy. The left column shows results for the $(8o, 3e)$ active space, and the right column shows results for the $(14o, 7e)$ active space.}
    \label{fig:multi_energy_grid}
\end{figure}